 \documentclass[final,5p,times,twocolumn]{elsarticle}

\usepackage{amssymb}

 \usepackage{lineno}

\biboptions{sort&compress}

\journal{arXiv}

\begin{document}

\begin{frontmatter}



\title{Concerning the Phase of the Time-Variation in the $^{36}$Cl Decay Rate}


\author[phy]{E. Fischbach\corref{cor}}
\address[phy]{Department of Physics, Purdue University, West Lafayette, IN 47907, USA}
\cortext[cor]{Corresponding author}
\ead{ephraim@purdue.edu}
\author[phy]{Tom Gruenwald}
\author[412]{D. Javorsek, II}
\address[412]{412th Test Wing, Edwards AFB, CA 93524, USA}
\author[NE,phy]{J. H. Jenkins}
\address[NE]{School of Nuclear Engineering, Purdue University, West Lafayette, IN 47907, USA}
\author[USAFA]{Robert H. Lee}
\address[USAFA]{Physics Department; U.S. Air Force Academy; 2354 Fairchild Dr., USAFA, CO 80840 USA}
\author[stan]{P. A. Sturrock}
\address[stan]{Center for Space Science and Astrophysics, Stanford University, Stanford, CA 94305, USA}

%

\begin{keyword}
Radioactivity \sep Beta Decay \sep Sun \sep Neutrinos

\end{keyword}

\end{frontmatter}



In a recent posting to the arXiv, Norman \cite{nor12} raises an interesting question relating to the
phase of the annually varying $^{36}$Cl measured decay rate as reported by two independent groups \cite{alb86,jen12}.
He correctly notes that the apparent phases reported in \cite{alb86,jen12} are not identical,
as might be expected in a model in which the annual decay-rate variation
is attributed simply to the varying Earth-Sun distance $R$. These determined phases are discussed in \citet{jav10} for the  \citet{alb86} data, and in \citet{jen12} for the second data set.  (By convention the phase
of the annual variation is the calendar day on which the decay rate is a maximum.)
In this note we address the question raised by Norman \cite{nor12}.

If the Sun were a uniform, homogeneous sphere producing energy and emitting particles
(e.g. neutrinos) at a constant, uniform rate, and the observed periodicities were due solely to the eccentricity of the Earth's orbit around the Sun, then the expected phase of decay data
would be either perihelion ($\sim$January 4) or aphelion ($\sim$July 4) depending on
the (as yet unknown) dynamics of the decay progress.  However, most of the
nuclides for which measured decay data are currently available exhibit a phase closer to
mid-February, rather than January 4.  Hence our first task is to understand
the origin of the mid-February phase.  In Ref. \cite{stu11} we propose that this
phase arises from a combination of two annually varying effects: the $1/R^2$
variation arising from the ellipticity of the Earth's orbit around the Sun,
and a North-South (latitudinal) asymmetry in neutrino production or propagation occurring in the Sun itself, for which there is
considerable independent evidence \citep{mas96,stu98,dor00,gon00,riv00,zha05}. This phase shift from perihelion has been seen in the phase determinations of two major solar neutrino observatories, as described in Refs. \citep{smy04,aha05,hos06,ran07}.

As we note in Ref.\cite{stu11} the North-South asymmetry 
effect alone would yield a phase $\sim$March 10 (or September 10) due to
the $7^\circ$ tilt of the solar axis of rotation relative to the ecliptic.
In this picture the mid-February phase would then result by combining the
$1/R^2$ effect ($\sim$ January 4) and the North-South asymmetry (March 10) with appropriate relative weights.  Since any North-South asymmetry would be
expected to be a variable feature on the Sun, varying during the $\sim$11 year
solar cycle, this could account for the variation in phase in
$^{36}$Cl between the BNL data set (from 1982-1990) and the OSURR data set
(2005-2011).

As we discuss elsewhere, there are other periodicities present in various decay data that we have examined, including a rotational signal at $\sim$32 days \citep{stu10a,stu10b}, and a Rieger-like periodicity at 2.11 yr$^{-1}$ \citep{stu11a}, which can be attributed to the Sun. All of these also exhibit variable features, which should not be surprising since the Sun is known to be a very variable star.

\section*{Acknowledgments}

The authors wish to thank Eric Norman, Rob de Meijer and Stuart Clark
for helpful correspondence on the phase question.






\begin{thebibliography}{18}
\expandafter\ifx\csname natexlab\endcsname\relax\def\natexlab#1{#1}\fi
\providecommand{\bibinfo}[2]{#2}
\ifx\xfnm\relax \def\xfnm[#1]{\unskip,\space#1}\fi
\bibitem[{Norman(2012)}]{nor12}
\bibinfo{author}{E.~B. Norman},
\newblock \bibinfo{title}{Additional experimental evidence against a solar
  influence on nuclear decay rates},
\newblock \bibinfo{journal}{ArXiv} \bibinfo{volume}{arXiv:1208.4357 [nucl-ex]} (\bibinfo{year}{2012})
  \bibinfo{pages}{1--1}.
\bibitem[{Alburger et~al.(1986)Alburger, Harbottle, and Norton}]{alb86}
\bibinfo{author}{D.~E. Alburger}, \bibinfo{author}{G.~Harbottle},
  \bibinfo{author}{E.~F. Norton},
\newblock \bibinfo{title}{Half-life of $^{32}$Si},
\newblock \bibinfo{journal}{Earth and Planetary Science Letters}
  \bibinfo{volume}{78} (\bibinfo{year}{1986}) \bibinfo{pages}{168--76}.
\bibitem[{Jenkins et~al.(2012)Jenkins, Herminghuysen, Blue, Fischbach,
  Javorsek~II, Kauffman, Mundy, Sturrock, and Talnagi}]{jen12}
\bibinfo{author}{J.~H. Jenkins}, \bibinfo{author}{K.~R. Herminghuysen},
  \bibinfo{author}{T.~E. Blue}, \bibinfo{author}{E.~Fischbach},
  \bibinfo{author}{D.~Javorsek~II}, \bibinfo{author}{A.~C. Kauffman},
  \bibinfo{author}{D.~W. Mundy}, \bibinfo{author}{P.~A. Sturrock},
  \bibinfo{author}{J.~W. Talnagi},
\newblock \bibinfo{title}{Additional experimental evidence for a solar
  influence on nuclear decay rates},
\newblock \bibinfo{journal}{Astroparticle Physics} \bibinfo{volume}{37}
  (\bibinfo{year}{2012}) \bibinfo{pages}{81--88}.
\bibitem[{Javorsek~II et~al.(2010)Javorsek~II, Sturrock, Lasenby, Lasenby,
  Buncher, Fischbach, Gruenwald, Hoft, Horan, Jenkins, Kerford, Lee, Longman,
  Mattes, Morreale, Morris, Mudry, Newport, O'Keefe, Petrelli, Silver, Stewart,
  and Terry}]{jav10}
\bibinfo{author}{D.~Javorsek~II}, \bibinfo{author}{P.~A. Sturrock},
  \bibinfo{author}{R.~N. Lasenby}, \bibinfo{author}{A.~N. Lasenby},
  \bibinfo{author}{J.~B. Buncher}, \bibinfo{author}{E.~Fischbach},
  \bibinfo{author}{J.~T. Gruenwald}, \bibinfo{author}{A.~W. Hoft},
  \bibinfo{author}{T.~J. Horan}, \bibinfo{author}{J.~H. Jenkins},
  \bibinfo{author}{J.~L. Kerford}, \bibinfo{author}{R.~H. Lee},
  \bibinfo{author}{A.~Longman}, \bibinfo{author}{J.~J. Mattes},
  \bibinfo{author}{B.~L. Morreale}, \bibinfo{author}{D.~B. Morris},
  \bibinfo{author}{R.~N. Mudry}, \bibinfo{author}{J.~R. Newport},
  \bibinfo{author}{D.~O'Keefe}, \bibinfo{author}{M.~A. Petrelli},
  \bibinfo{author}{M.~A. Silver}, \bibinfo{author}{C.~A. Stewart},
  \bibinfo{author}{B.~Terry},
\newblock \bibinfo{title}{Power spectrum analyses of nuclear decay rates},
\newblock \bibinfo{journal}{Astroparticle Physics} \bibinfo{volume}{34}
  (\bibinfo{year}{2010}) \bibinfo{pages}{173--178}.
\bibitem[{Sturrock et~al.(2011)Sturrock, Buncher, Fischbach, Javorsek~II,
  Jenkins, and Mattes}]{stu11}
\bibinfo{author}{P.~Sturrock}, \bibinfo{author}{J.~Buncher},
  \bibinfo{author}{E.~Fischbach}, \bibinfo{author}{D.~Javorsek~II},
  \bibinfo{author}{J.~Jenkins}, \bibinfo{author}{J.~Mattes},
\newblock \bibinfo{title}{Concerning the phases of the annual variations of
  nuclear decay rates},
\newblock \bibinfo{journal}{The Astrophysical Journal} \bibinfo{volume}{737}
  (\bibinfo{year}{2011}) \bibinfo{pages}{65}.
\bibitem[{Massetti and Storini(1996)}]{mas96}
\bibinfo{author}{S.~Massetti}, \bibinfo{author}{M.~Storini},
\newblock \bibinfo{title}{Spacetime modulation of solar neutrino flux:
  1970-1992},
\newblock \bibinfo{journal}{Astrophysical Journal} \bibinfo{volume}{472}
  (\bibinfo{year}{1996}) \bibinfo{pages}{827--31}.
\bibitem[{Sturrock et~al.(1998)Sturrock, Walther, and Wheatland}]{stu98}
\bibinfo{author}{P.~A. Sturrock}, \bibinfo{author}{G.~Walther},
  \bibinfo{author}{M.~S. Wheatland},
\newblock \bibinfo{title}{Apparent latitudinal modulation of the solar neutrino
  flux},
\newblock \bibinfo{journal}{Astrophysical Journal} \bibinfo{volume}{507}
  (\bibinfo{year}{1998}) \bibinfo{pages}{978--83}.
\bibitem[{Dorman(2000)}]{dor00}
\bibinfo{author}{L.~I. Dorman},
\newblock \bibinfo{title}{The asymmetry of solar-neutrino fluxes},
\newblock \bibinfo{journal}{Physics of Atomic Nuclei} \bibinfo{volume}{63}
  (\bibinfo{year}{2000}) \bibinfo{pages}{989--92}.
\bibitem[{Gonzalez-Garcia et~al.(2000)Gonzalez-Garcia, de~Holanda, and
  Pena-Garay}]{gon00}
\bibinfo{author}{M.~C. Gonzalez-Garcia}, \bibinfo{author}{P.~C. de~Holanda},
  \bibinfo{author}{C.~Pena-Garay},
\newblock \bibinfo{title}{Seasonal dependence in the solar neutrino flux},
\newblock volume~\bibinfo{volume}{81} of \textit{\bibinfo{series}{Nucl. Phys.
  B, Proc. Suppl. (Netherlands)}}, \bibinfo{publisher}{Elsevier},
  \bibinfo{address}{Netherlands}, \bibinfo{year}{2000}, pp.
  \bibinfo{pages}{89--94}.
\bibitem[{Rivin and Obridko(2000)}]{riv00}
\bibinfo{author}{Y.~R. Rivin}, \bibinfo{author}{V.~N. Obridko},
\newblock \bibinfo{title}{Seasonal variations in solar high-energy neutrino
  flux and their probable source},
\newblock \bibinfo{journal}{Solar System Research} \bibinfo{volume}{34}
  (\bibinfo{year}{2000}) \bibinfo{pages}{501--8}.
\bibitem[{Zhao et~al.(2005)Zhao, Hoeksema, and Scherrer}]{zha05}
\bibinfo{author}{X.~P. Zhao}, \bibinfo{author}{J.~T. Hoeksema},
  \bibinfo{author}{P.~H. Scherrer},
\newblock \bibinfo{title}{Prediction and understanding of the north-south
  displacement of the heliospheric current sheet},
\newblock \bibinfo{journal}{Journal of Geophysical Research-Part A-Space
  Physics} \bibinfo{volume}{110} (\bibinfo{year}{2005}) \bibinfo{pages}{8 pp.}
\bibitem[{Smy et~al.(2004)Smy, Ashie, Fukuda, and et~al.}]{smy04}
\bibinfo{author}{M.~B. Smy}, \bibinfo{author}{Y.~Ashie},
  \bibinfo{author}{S.~Fukuda}, \bibinfo{author}{et~al.},
\newblock \bibinfo{title}{Precise measurement of the solar neutrino day-night
  and seasonal variation in Super-Kamiokande-I},
\newblock \bibinfo{journal}{Physical Review D} \bibinfo{volume}{69}
  (\bibinfo{year}{2004}) \bibinfo{pages}{11104--1}.
\bibitem[{Aharmim et~al.(2005)Aharmim, Ahmed, Anthony, and et~al.}]{aha05}
\bibinfo{author}{B.~Aharmim}, \bibinfo{author}{S.~N. Ahmed},
  \bibinfo{author}{A.~E. Anthony}, \bibinfo{author}{et~al.},
\newblock \bibinfo{title}{Search for periodicities in the $^{8}$B solar
  neutrino flux measured by the Sudbury Neutrino Observatory},
\newblock \bibinfo{journal}{Physical Review D} \bibinfo{volume}{72}
  (\bibinfo{year}{2005}) \bibinfo{pages}{052010}.
\bibitem[{Hosaka et~al.(2006)Hosaka, Ishihara, Kameda, Koshio, and
  et~al.}]{hos06}
\bibinfo{author}{J.~Hosaka}, \bibinfo{author}{K.~Ishihara},
  \bibinfo{author}{J.~Kameda}, \bibinfo{author}{Y.~Koshio},
  \bibinfo{author}{et~al.},
\newblock \bibinfo{title}{Solar neutrino measurements in Super-Kamiokande-I},
\newblock \bibinfo{journal}{Physical Review D (Particles and Fields)}
  \bibinfo{volume}{73} (\bibinfo{year}{2006}) \bibinfo{pages}{112001--33}.
\bibitem[{Ranucci and Sello(2007)}]{ran07}
\bibinfo{author}{G.~Ranucci}, \bibinfo{author}{S.~Sello},
\newblock \bibinfo{title}{Search for periodicities in the experimental solar
  neutrino data: a wavelet approach},
\newblock \bibinfo{journal}{Physical Review D} \bibinfo{volume}{75}
  (\bibinfo{year}{2007}) \bibinfo{pages}{73011--1}.
\bibitem[{Sturrock et~al.(2010{\natexlab{a}})Sturrock, Buncher, Fischbach,
  Gruenwald, Javorsek~II, Jenkins, Lee, Mattes, and Newport}]{stu10a}
\bibinfo{author}{P.~A. Sturrock}, \bibinfo{author}{J.~B. Buncher},
  \bibinfo{author}{E.~Fischbach}, \bibinfo{author}{J.~T. Gruenwald},
  \bibinfo{author}{D.~Javorsek~Ii}, \bibinfo{author}{J.~H. Jenkins},
  \bibinfo{author}{R.~H. Lee}, \bibinfo{author}{J.~J. Mattes},
  \bibinfo{author}{J.~R. Newport},
\newblock \bibinfo{title}{Power spectrum analysis of BNL decay rate data},
\newblock \bibinfo{journal}{Astroparticle Physics} \bibinfo{volume}{34}
  (\bibinfo{year}{2010}{\natexlab{a}}) \bibinfo{pages}{121--127}.
\bibitem[{Sturrock et~al.(2010{\natexlab{b}})Sturrock, Buncher, Fischbach,
  Gruenwald, Javorsek, Jenkins, Lee, Mattes, and Newport}]{stu10b}
\bibinfo{author}{P.~A. Sturrock}, \bibinfo{author}{J.~Buncher},
  \bibinfo{author}{E.~Fischbach}, \bibinfo{author}{J.~Gruenwald},
  \bibinfo{author}{D.~Javorsek}, \bibinfo{author}{J.~Jenkins},
  \bibinfo{author}{R.~Lee}, \bibinfo{author}{J.~Mattes},
  \bibinfo{author}{J.~Newport},
\newblock \bibinfo{title}{Power spectrum analysis of Physikalisch-Technische
  Bundesanstalt decay-rate data: Evidence for solar rotational modulation},
\newblock \bibinfo{journal}{Solar Physics} \bibinfo{volume}{267}
  (\bibinfo{year}{2010}{\natexlab{b}}) \bibinfo{pages}{251--265}.
\bibitem[{Sturrock et~al.(2011)Sturrock, Fischbach, and Jenkins}]{stu11a}
\bibinfo{author}{P.~A. Sturrock}, \bibinfo{author}{E.~Fischbach},
  \bibinfo{author}{J.~H. Jenkins},
\newblock \bibinfo{title}{Further evidence suggestive of a solar influence on
  nuclear decay rates},
\newblock \bibinfo{journal}{Solar Physics} \bibinfo{volume}{272}
  (\bibinfo{year}{2011}) \bibinfo{pages}{1--10}.

\end{thebibliography}



\end{document}